\documentclass[aps,prl,showpacs,twocolumn,groupedaddress]{revtex4}
\usepackage{graphicx}
\usepackage{dcolumn}
\usepackage{bm}
\usepackage{psfrag}
\usepackage{subfigure}
\usepackage{amsfonts, amssymb, latexsym}
\usepackage{times}
\newcommand{\tc}{,~}
\newcommand{\fnz}{\footnotesize}

\newcommand{\ket}[1]{\left| #1 \right\rangle}

\newcommand{\ua}{\uparrow}
\newcommand{\da}{\downarrow}

\newcommand{\Ua}{\Uparrow}

\newcommand{\Ra}{\Rightarrow}

\newcommand{\q}{\mathrm}

\newcommand{\duratio}{\mathit{R^{du}}}

\newcommand{\ncaltech}{1}
\newcommand{\ncalsu}{2}
\newcommand{\ncf}{3}
\newcommand{\nfiu}{4}
\newcommand{\nfsu}{5}
\newcommand{\nuiuc}{6}
\newcommand{\ninfn}{7}
\newcommand{\njlab}{8}
\newcommand{\nksu}{9}
\newcommand{\nkentucky}{10}
\newcommand{\numd}{11}
\newcommand{\numass}{12}
\newcommand{\nmit}{13}
\newcommand{\nunh}{14}
\newcommand{\nodu}{15}
\newcommand{\nrutgers}{16}
\newcommand{\nsaclay}{17}
\newcommand{\nsyracuse}{18}
\newcommand{\ntelaviv}{19}
\newcommand{\ntemple}{20}
\newcommand{\nuva}{21}
\newcommand{\nwm}{22}

\begin{document}

\preprint{APS/123-QED}


\title{Precision Measurement of the Neutron Spin Asymmetry $A_1^n$ and Spin-Flavor 
Decomposition in the Valence Quark Region}

\author{
X.~Zheng$^{\nmit}$,
K.~Aniol$^{\ncalsu}$,
D.~S.~Armstrong$^{\nwm}$,
T.~D.~Averett$^{{\njlab,\nwm}}$,
W.~Bertozzi$^{\nmit}$,
S.~Binet$^{\nuva}$,
E.~Burtin$^{\nsaclay}$,
E.~Busato$^{\nrutgers}$, 
C.~Butuceanu$^{\nwm}$,
J.~Calarco$^{\nunh}$,
A.~Camsonne$^{\ncf}$,
G.~D.~Cates$^{\nuva}$,
Z.~Chai$^{\nmit}$,
J.-P.~Chen$^{\njlab}$,
Seonho~Choi$^{\ntemple}$,
E.~Chudakov$^{\njlab}$,
F.~Cusanno$^{\ninfn}$,
R.~De~Leo$^{\ninfn}$,
A.~Deur$^{\nuva}$,
S.~Dieterich$^{\nrutgers}$, 
D.~Dutta$^{\nmit}$,
J.~M.~Finn$^{\nwm}$,
S.~Frullani$^{\ninfn}$,
H.~Gao$^{\nmit}$, 
J.~Gao$^{\ncaltech}$,
F.~Garibaldi$^{\ninfn}$,
S.~Gilad$^{\nmit}$,
R.~Gilman$^{{\njlab,\nrutgers}}$, 
J.~Gomez$^{\njlab}$,
J.-O.~Hansen$^{\njlab}$,
D.~W.~Higinbotham$^{\nmit}$, 
W.~Hinton$^{\nodu}$,
T.~Horn$^{\numd}$,
C.W.~de~Jager$^{\njlab}$,
X.~Jiang$^{\nrutgers}$, 
L.~Kaufman$^{\numass}$, 
J.~Kelly$^{\numd}$, 
W.~Korsch$^{\nkentucky}$,
K.~Kramer$^{\nwm}$,
J.~LeRose$^{\njlab}$,
D.~Lhuillier$^{\nsaclay}$,
N.~Liyanage$^{\njlab}$,
D.J.~Margaziotis$^{\ncalsu}$,
F.~Marie$^{\nsaclay}$,
P.~Markowitz$^{\nfiu}$,
K.~McCormick$^{\nksu}$,
Z.-E.~Meziani$^{\ntemple}$,
R.~Michaels$^{\njlab}$,
B.~Moffit$^{\nwm}$,
S.~Nanda$^{\njlab}$,
D.~Neyret$^{\nsaclay}$,
S.~K.~Phillips$^{\nwm}$, 
A.~Powell$^{\nwm}$,
T.~Pussieux$^{\nsaclay}$,
B.~Reitz$^{\njlab}$,
J.~Roche$^{\nwm}$, 
R.~Roche$^{\nfsu}$,
M.~Roedelbronn$^{\nuiuc}$,
G.~Ron$^{\ntelaviv}$,
M.~Rvachev$^{\nmit}$, 
A.~Saha$^{\njlab}$,
N.~Savvinov$^{\numd}$,
J.~Singh$^{\nuva}$,
S.~\v{S}irca$^{\nmit}$,
K.~Slifer$^{\ntemple}$,
P.~Solvignon$^{\ntemple}$,
P.~Souder$^{\nsyracuse}$,
D.J.~Steiner$^{\nwm}$, 
S.~Strauch$^{\nrutgers}$,
V.~Sulkosky$^{\nwm}$, 
A.~Tobias$^{\nuva}$,
G.~Urciuoli$^{\ninfn}$,
A.~Vacheret$^{\numass}$,
B.~Wojtsekhowski$^{\njlab}$,
H.~Xiang$^{\nmit}$, 
Y.~Xiao$^{\nmit}$, 
F.~Xiong$^{\nmit}$, 
B.~Zhang$^{\nmit}$, 
L.~Zhu$^{\nmit}$,
X.~Zhu$^{\nwm}$,
P.A.\.Zo{\l}nierczuk$^{\nkentucky}$,
}
\address{
\baselineskip 2 pt
\vskip 0.3 cm
{\rm The Jefferson Lab Hall A Collaboration} \break
\vskip 0.1 cm
{\small{
{$^{\ncaltech}$California Institute of Technology, Pasadena, CA 91125} \break
{$^{\ncalsu}$California State University, Los Angeles, Los Angeles, CA 90032} \break
{$^{\ncf}$Universit\'{e} Blaise Pascal Clermont-Ferrand et CNRS/IN2P3 LPC 63\tc 177 Aubi\`{e}re Cedex, France} \break
{$^{\nfiu}$Florida International University, Miami, FL 33199} \break
{$^{\nfsu}$Florida State University, Tallahassee, FL 32306} \break
{$^{\nuiuc}$University of Illinois, Urbana, IL 61801} \break
{$^{\ninfn}$Istituto Nazionale di Fisica Nucleare\tc Sezione Sanit\`a\tc 00161 Roma, Italy} \break
{$^{\njlab}$Thomas Jefferson National Accelerator Facility\tc Newport News, VA 23606} \break
{$^{\nksu}$Kent State University, Kent, OH 44242} \break
{$^{\nkentucky}$University of Kentucky, Lexington, KY 40506} \break
{$^{\numd}$University of Maryland, College Park, MD 20742} \break
{$^{\numass}$University of Massachusetts Amherst\tc Amherst, MA 01003} \break
{$^{\nmit}$Massachusetts Institute of Technology\tc Cambridge, MA 02139} \break
{$^{\nunh}$University of New Hampshire, Durham, NH 03824} \break
{$^{\nodu}$Old Dominion University, Norfolk, VA 23529} \break
{$^{\nrutgers}$Rutgers, The State University of New Jersey\tc Piscataway, NJ 08855} \break
{$^{\nsaclay}$CEA Saclay, DAPNIA/SPhN\tc F-91191 Gif sur Yvette, France} \break
{$^{\nsyracuse}$Syracuse University, Syracuse, NY 13244} \break
{$^{\ntelaviv}$University of Tel Aviv, Tel Aviv 69978, Israel} \break
{$^{\ntemple}$Temple University, Philadelphia, PA 19122} \break
{$^{\nuva}$University of Virginia, Charlottesville, VA 22904} \break
{$^{\nwm}$College of William and Mary\tc Williamsburg, VA 23187} \break
}}
}
\begin{abstract}                
We have measured the neutron spin asymmetry $A_1^n$ with  
high precision at three kinematics in the deep inelastic  
region at $x=0.33$, $0.47$ and $0.60$, and $Q^2=2.7$,     
$3.5$ and $4.8$~(GeV/c)$^2$, respectively. Our results    
unambiguously show, for the first time, that $A_1^n$      
crosses zero around $x=0.47$ and becomes significantly    
positive at $x=0.60$.  Combined with the world            
proton data, polarized quark distributions   
were extracted. Our results, in general, agree            
with relativistic constituent quark models and with perturbative       
quantum chromodynamics (pQCD) analyzes based on the       
earlier data. However they deviate from pQCD              
predictions based on hadron helicity conservation.        
%

\end{abstract}

\maketitle

After over twenty-five years of experiments measuring nucleon spin 
structure, it is now widely accepted that 
the intrinsic quark spin contributes only a small fraction (20\%-30\%) 
of the total nucleon spin.  The spin sum rule~\cite{theory:Nspinsr}
indicates that the remaining part is carried by the quarks and gluons 
orbital angular momentum (OAM) and gluon spin. 

Here we present precise data in a new kinematic region
where the Bjorken scaling variable $x$ is large.  For these
kinematics, the valence quarks dominate and ratios of structure
functions can be estimated based on our knowledge of the interactions
between quarks.  Specifically, in the limit of large $Q^2$
(the four momentum transfer squared), the 
asymmetry $A_1$  (the ratio of the polarized and the unpolarized
structure functions $g_1/F_1$) is expected 
to approach $1$ as $x\rightarrow 1$. This is a dramatic prediction, 
since all previous data on the neutron $A_1^n$ are either negative 
or consistent with zero.  Furthermore, in the region $x>0.3$, both 
sea-quark and gluon contributions are small and the physics of the valence 
quarks can be exposed. 
Relativistic constituent quark models (RCQM, which include OAM) and
leading-order pQCD predictions assuming hadron-helicity-conservation
(no OAM) make dramatically different predictions for the proton
down-quark polarized distribution in the valence quark region. A more complete QCD
calculation, describing OAM at the current-quark and gluon level,
might agree with the RCQM description. The connection 
between these descriptions is of paramount importance to a 
complete description of the nucleon spin using QCD. Thus, precision data 
in the valence quark region are crucial to improve our understanding of 
the nucleon spin.

$A_1$ is known as the nucleon virtual-photon asymmetry and is extracted 
from the polarized deep inelastic scattering (DIS) cross sections as
$A_1 = ({\sigma_{1/2}-\sigma_{3/2}})/({\sigma_{1/2}+\sigma_{3/2}})$,
where $\sigma_{1/2\,(3/2)}$ is the total virtual photo-absorption cross 
section for the nucleon with a projection of $1/2$ ($3/2$) for the total 
spin along the direction of photon momentum~\cite{book:ph-had-in}.  
At finite $Q^2$, $A_1$ is related to the polarized and unpolarized 
structure functions $g_1$, $g_2$ and $F_1$ through
\begin{eqnarray}
 A_1(x,Q^2) &=& \big[{g_1(x,Q^2)-\gamma^2g_2(x,Q^2)}\big]/{F_1(x,Q^2)}~,~~
\end{eqnarray}
where $\gamma^2={4M^2 x^2}/{Q^2}$, $M$ is the 
nucleon mass, $Q^2=4EE^\prime\sin^2{(\theta/2)}$, $x={Q^2}/({2M\nu})$, 
$E$ is the beam energy, $E^\prime$ is the energy of the scattered electron,
$\nu=E-E^\prime$ is the energy transfer to the target and 
$\theta$ is the scattering angle in the lab frame.  
At high $Q^2$, one has $\gamma^2 \ll 1$ and $A_1\approx g_1/F_1$. Since 
$g_1$ and $F_1$ follow roughly the same $Q^2$ evolution in leading order 
QCD, $A_1$ is expected to vary quite slowly with $Q^2$.

To first approximation, the constituent quarks in the neutron can be 
described by an SU(6) symmetric wave-function~\cite{theory:su6close}
\begin{eqnarray}\label{equ:nwf}
  \ket{n\ua} = \frac{1}{\sqrt{2}}\ket{d^\ua (du)_{0,0,0}}
   +\frac{1}{\sqrt{18}}\ket{d^\ua (du)_{1,1,0}}~~~~~~~~~~~~~~~~~~&&  \\
     -\frac{1}{3} \ket{d^\da (du)_{1,1,1}}
     -\frac{1}{3}\ket{u^\ua (dd)_{1,1,0}}
     +\frac{\sqrt{2}}{3}\ket{u^\da (dd)_{1,1,1}},&&\nonumber
\end{eqnarray}
where $u$ ($d$) is the wavefunction of up (down) quark inside the neutron 
and the subscripts refer to $I$, $S$ and $S_z$, the total isospin, 
total spin and the spin projection of the spectator diquark state.
In this limit both $S=1$ and $S=0$ 
diquark states contribute equally to the observables of interest, leading 
to the predictions of $A_1^p={5}/{9}$ and $A_1^n=0$.

However, from measurements of the $x$-dependence of the ratio $F_2^p/F_2^n$ 
in unpolarized DIS~\cite{theory:su6breaking} it is known that the SU(6) 
symmetry is broken.  A phenomenological SU(6) symmetry breaking mechanism 
is the hyperfine interaction among the quarks. Its effect on the nucleon 
wave-function is to lower the energy of the $S=0$ diquark state, allowing 
the first term of Eq.~(\ref{equ:nwf})
to be more 
stable and hence to dominate the high momentum tail of the quark distributions, 
which is probed as $x\to 1$.  In this picture one obtains $\Delta u/u\to 1$, 
$\Delta d/d\to -1/3$ and ${A}_{1}^{n,p}\to 1$ as $x\to 1$, with $\Delta u$($\Delta d$) 
and $u$($d$) the polarized and unpolarized quark distributions for the 
$u$($d$) quark in the proton.  The hyperfine interaction is often used to break SU(6) 
symmetry in RCQM to calculate 
$A_1^n(x)$ and $A_1^p(x)$ in the region 
$0.4<x<1$~\cite{theory:cqm_ze,theory:cqm,theory:ma}. 

In the pQCD approach~\cite{theory:farrar,theory:farrar2} it was noted that 
the quark-gluon interactions cause only the $S=1$, $S_z=1$ diquark states 
to be suppressed as $x\to 1$, rather than the full $S=1$ states as in the 
case for the hyperfine interaction.  By assuming zero quark OAM and helicity 
conservation, it has been shown further that a quark with $x\to 1$ must have 
the same helicity as the 
nucleon.  This mechanism has been referred to as hadron helicity conservation (HHC)
and was used to build parton distribution functions~\cite{theory:bbs} and to fit DIS 
data~\cite{theory:lssbbs}.
In this approach one has ${A}_{1}^{n,p}\to 1$, $\Delta u/u\to 1$ and $\Delta d/d\to 1$ 
as $x\to 1$. This is one of the few places where QCD can make a prediction for the 
structure function ratios. 

The HHC is based on leading order pQCD where the quark OAM is assumed to 
be zero. Recent data on the 
tensor polarization in elastic $e-^2$H scattering~\cite{data:cebaf-t20},
neutral pion photo-production~\cite{data:cebaf-gammap}
and the proton form factors~\cite{data:cebaf-F2p,data:cebaf-Gep}
are in disagreement with HHC predictions.  It has been suggested that 
effects beyond leading-order pQCD, such as the quark 
OAM~\cite{theory:miller,theory:transpdf,theory:ji},
might play an important role in processes involving spin flips.  
Calculations including quark OAM were performed to interpret 
the proton form factor data~\cite{theory:ji}.
These kinds of calculations may be possible in the future for $A_1^n$ and 
other observables in the large $x$ region~\cite{ji_pv}.

Other available predictions for $A_1^n$ include those from the bag 
model~\cite{theory:bag}, the LSS Next-to-Leading Order (NLO) polarized parton 
densities~\cite{theory:lss2001}, the chiral soliton model~\cite{theory:chi},
a global NLO QCD analysis of DIS data based on a 
statistical picture of the nucleon~\cite{theory:stat}, and quark-hadron duality 
based on three different SU(6) symmetry breaking scenarios~\cite{theory:duality}. 

We measured inclusive deep inelastic scattering of longitudinally polarized electrons 
from a polarized $^3$He target in Hall~A of the Thomas Jefferson National Accelerator 
Facility.  Data were collected at three kinematics, $x=0.33$, $0.47$ 
and $0.60$, with $Q^2=2.7$, $3.5$ and $4.8$~(GeV/c)$^2$, respectively. The invariant 
mass squared $W^2=M^2+2M\nu-Q^2$ was above the resonance region. The 
parallel ($A_\parallel$) and perpendicular ($A_\perp$) asymmetries were measured. 
They are defined as
\begin{eqnarray}
 A_\parallel &=& \frac{\sigma^{\da\Ua}-\sigma^{\ua\Ua}}{\sigma^{\da\Ua}+\sigma^{\ua\Ua}}~~
\q{and}~~A_\perp = \frac{\sigma^{\da\Ra}-\sigma^{\ua\Ra}}{\sigma^{\da\Ra}+\sigma^{\ua\Ra}}~,
\end{eqnarray}
where $\sigma^{\da\Ua}$ ($\sigma^{\ua\Ua}$) is the cross section for a longitudinally 
(with respect to the beamline) 
polarized target with the electron spin aligned antiparallel (parallel) to the target spin; 
$\sigma^{\da\Ra}$ ($\sigma^{\ua\Ra}$) is the cross section for a transversely polarized 
target with the electron spin aligned antiparallel (parallel) to the beam direction, and 
with the scattered electrons detected on the same side of the beamline as that to which 
the target spin is pointing. One can extract $A_1$ as 
\begin{eqnarray}\label{equ:a1apap}
  {A_1} &=& \frac{A_{\parallel}}{D(1+\eta\xi)} - \frac{\eta A_\perp}{d(1+\eta\xi)}~,
\end{eqnarray}
where $D=({1-\epsilon E^\prime/E})/({1+\epsilon R})$, $d=D\sqrt{2\epsilon/(1+\epsilon)}$, 
$\eta=\epsilon\sqrt{Q^2}/(E-E^\prime\epsilon)$, $\xi=\eta(1+\epsilon)/(2\epsilon)$,
$\epsilon = 1/[1+2(1+1/\gamma^2)\tan^2(\theta/2)]$ and $R$ is the ratio of the longitudinal
and transverse virtual photon absorption cross sections
$\sigma_L/\sigma_T$~\cite{book:ph-had-in}.  Similarly, the ratio of structure 
functions is given by ${g_1}/{F_1} = [{A_\parallel+A_\perp\tan(\theta/2)}]/{D^\prime}$,
with $D^\prime=[{(1-\epsilon)(2-y)}]/[y(1+\epsilon R)]$ and $y=\nu/E$.

The polarized electron beam was produced by illuminating a strained 
GaAs photocathode with circularly polarized light. We used a beam energy of $5.7$~GeV.
The beam polarization of $P_b=(79.7\pm 2.4)\%$
was measured regularly by M$\o$ller polarimetry and was monitored by Compton polarimetry.
The beam helicity 
was flipped at a frequency of $30$~Hz. To reduce possible systematic errors, 
data were taken for four 
different beam helicity and target polarization configurations for the parallel setting and 
two for the perpendicular setting.

The polarized $^3$He target is based on the principles of optical pumping and spin 
exchange. The target cell is a $25$~cm long glass vessel.
The in-beam target density was about $3.5\times 10^{20}$~$^3$He/cm$^3$.
The target polarization was measured by both
the NMR technique of adiabatic fast passage~\cite{exp:AFP}, and a technique
based on electron paramagnetic resonance~\cite{exp:EPR}.
The average in-beam target polarization was $P_t=(40\pm 1.5)\%$ 
at a typical beam current of $12$~$\mu$A. 
The product of the beam and target polarizations was verified at the level of 
$\Delta(P_bP_t)/(P_bP_t)\leqslant~4.5\%$ by measuring the longitudinal asymmetry of
$\vec e-^3\overrightarrow{\q{He}}$ elastic scattering.

The scattered electrons were detected by the Hall A High Resolution Spectrometer
(HRS) pair~\cite{exp:NIM} at two scattering angles of $35^\circ$ and $45^\circ$.  
A CO$_2$ gas \v{C}erenkov detector and a double-layered lead-glass 
shower counter were used to separate electrons from the pion background.  
The combined pion rejection factor provided by the two detectors was found to be better 
than $10^{4}$ for both HRSs, with a $99\%$ identification efficiency for electrons.

The asymmetries are extracted from the data as
 $A_{\parallel,\perp}={A_{raw}}/{(fP_bP_t)}+\Delta A_{\parallel,\perp}^{RC}$,
where $A_{raw}$ is the raw asymmetry and $f=0.92\sim 0.94$ is the target dilution 
factor due to a small amount of unpolarized N$_2$ mixed with the polarized $^3$He gas.
Radiative corrections $\Delta A_{\parallel,\perp}^{RC}$ were performed
for both the internal and the external radiation effects. Internal radiative corrections 
were applied using POLRAD2.0~\cite{ana:polrad},
the most up-to-date structure functions and our data for the neutron polarized 
structure functions.  External radiative corrections were performed based on the 
procedure first described by Mo and Tsai~\cite{ana:motsai}. The uncertainty in the 
correction was studied by using various fits~\cite{thesis:zheng} to the world data 
for $F_2$, $g_1$, $g_2$ and $R$.
False asymmetries were checked to be negligible by measuring the asymmetries of 
polarized $e^-$ beam scattering off an unpolarized $^{12}$C target. 

From $A_{\parallel,\perp}$ one can calculate $A_1^{^3\q{He}}$ using Eq.~(\ref{equ:a1apap}). 
A $^3$He model which includes $S$, $S^\prime$, $D$ states and pre-existing $\Delta(1232)$ 
component in the $^3$He wavefunction~\cite{theory:3Hecmplt} was used 
for extracting $A_1^n$ from $A_1^{^3\q{He}}$.  It gives
\begin{eqnarray}
 A_1^n&=&\frac{F_2^{^3\q{He}}
	[{A_1^{^3\q{He}}}-2\frac{F_2^p}{F_2^{^3\q{He}}}P_pA_1^p(1-\frac{0.014}{2P_p})]}
	{P_nF_2^n(1+\frac{0.056}{P_n})}~, \label{equ:he3ton}
\end{eqnarray}
where $P_n=0.86^{+0.036}_{-0.02}$ and $P_p=-0.028^{+0.009}_{-0.004}$ are the effective 
nucleon polarizations of the neutron and the proton inside 
$^3$He~\cite{theory:3Heconv,theory:3Hecmplt,theory:PnPpBonn}.
We used the latest world proton and deuteron fits~\cite{ana:f2nmc95,ana:r1998} for 
$F_2$ and $R$, with nuclear effects corrected~\cite{theory:EMC_wally}.
The $A_1^p$ contribution was obtained by fitting the world proton data~\cite{thesis:zheng}.
Compared to the convolution approach~\cite{theory:3Heconv} used by previous polarized $^3$He 
experiments, Eq.~(\ref{equ:he3ton}) increases the value of $A_1^n$ by
$0.01-0.02$ in the region $0.2<x<0.7$, which is small compared to our statistical error bars.
Eq.~(\ref{equ:he3ton}) was also used for extracting $g_1^n/F_1^n$ from $g_1^{^3\q{He}}/F_1^{^3\q{He}}$
by substituting $g_1/F_1$ for $A_1$. 

\begin{figure}[h]
 \begin{center}
 \includegraphics[angle = 0, width=218pt]{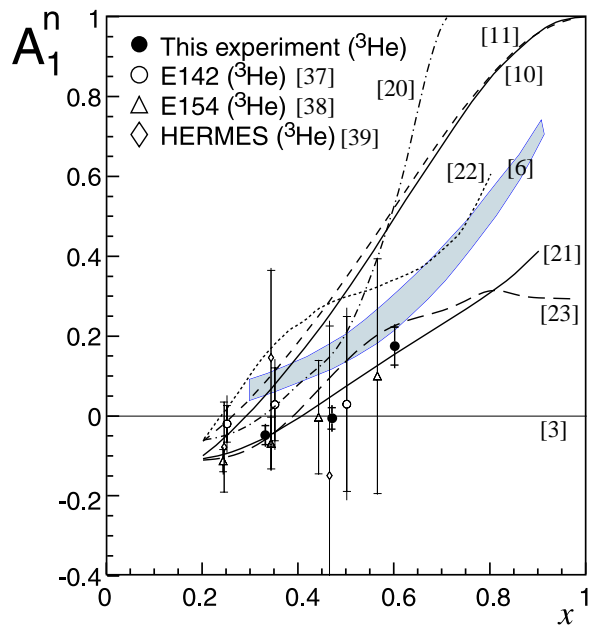}
 \put(-40.5,220.0){\cite{theory:lssbbs}}
 \put(-31.5,205.4){\cite{theory:bbs}}
 \put(-53.5,168.6){\cite{theory:chi}}
 \put(-30.6,170.1){\cite{theory:cqm}}
 \put(-17.1,137.4){\cite{theory:lss2001}}
 \put(-18,115.0){\cite{theory:stat}}
 \put(-80,200.0){\cite{theory:bag}}
 \put(-18,70.8){\cite{theory:su6close}}
 \put(-110.4,204.75){\cite{data:a1ng1n-e142}}
 \put(-110.4,193.05){\cite{data:a1ng1n-e154}}
 \put(-93.6,181.35){\cite{data:a1ng1n-hermes}}
 \end{center}
\caption{Our $A_1^n$ results compared with theoretical predictions and existing
data obtained from a polarized $^3$He 
target~\cite{data:a1ng1n-e142,data:a1ng1n-e154,data:a1ng1n-hermes}. 
Curves: 
predictions of $A_1^n$ from SU(6) symmetry (zero)~\cite{theory:su6close},
constituent quark model (shaded band)~\cite{theory:cqm} and
statistical model (long-dashed)~\cite{theory:stat};
predictions of ${g_1^n}/{F_1^n}$ from pQCD HHC based BBS parameterization (higher 
solid)~\cite{theory:bbs} and LSS(BBS) parameterization (dashed)~\cite{theory:lssbbs},
bag model with the effect
of hyperfine interaction but without meson cloud (dash-dotted)~\cite{theory:bag},
LSS~2001 NLO polarized parton densities 
(lower solid)~\cite{theory:lss2001}
and chiral soliton model (dotted)~\cite{theory:chi}.
}
\label{fig:result_a1n2}
\end{figure}

Results for $A_1^n$ and $g_1^n/F_1^n$ are given in Table~\ref{tab:result_a1n}. 
The $A_1^n$ results are shown in Fig.~\ref{fig:result_a1n2}.
The smaller and full error bars show the statistical and total errors, respectively. 
The largest systematic error comes from the uncertainties in $P_p$ and $P_n$.
\renewcommand{\arraystretch}{1.3}
\begin{table}[h]
\caption{Results for $A_1^n$ and $g_1^n/F_1^n$, $Q^2$ values are given in (GeV/c)$^2$, 
errors are given as $\pm$ statistical $\pm$ systematic.}
\label{tab:result_a1n}
\begin{ruledtabular}
\begin{tabular}{c|c|c|c}
 $x$  & $Q^2$ & $A_1^n$                 & $g_1^n/F_1^n$          \\ \hline 
$0.33$ & $2.71$ & $ -0.048\pm 0.024^{+ 0.015}_{-0.016}$ & $-0.043\pm 0.022^{+ 0.009}_{-0.009}$ \\
$0.47$ & $3.52$ & $ -0.006\pm 0.027^{+ 0.019}_{-0.019}$ & $+0.040\pm 0.035^{+ 0.011}_{-0.011}$ \\
$0.60$ & $4.83$ & $ +0.175\pm 0.048^{+ 0.026}_{-0.028}$ & $+0.124\pm 0.045^{+ 0.016}_{-0.017}$ \\
\end{tabular}
\end{ruledtabular}
\end{table}
\renewcommand{\arraystretch}{1.0}

The new datum at $x=0.33$ is in good agreement with world data.  
For $x>0.4$, the precision of $A_1^n$ data has been improved by about an order 
of magnitude.  This is the first experimental evidence that $A_1^n$ becomes 
positive at large~$x$. 
Among all model-based 
calculations~\cite{theory:su6close,theory:cqm,theory:bbs,theory:lssbbs,theory:bag,theory:chi}, 
the trend of our data is consistent with the RCQM predictions~\cite{theory:cqm} 
which suggest that $A_1^n$ becomes increasingly positive at even higher $x$. 
However they do not agree with the BBS~\cite{theory:bbs} and LSS(BBS)~\cite{theory:lssbbs} 
parameterizations in which HHC is imposed. Our data are in good agreement
with the LSS 2001 pQCD fit to previous data~\cite{theory:lss2001} and a global NLO QCD
analysis of DIS data using a statistical picture of the nucleon~\cite{theory:stat}.

Assuming the strange quark distributions $s(x)$, $\bar{s}(x)$, $\Delta s(x)$ and 
$\Delta \bar{s}(x)$ to be negligible in the region $x>0.3$, and ignoring any $Q^2$ 
dependence, one can extract polarized quark distribution functions based on the 
quark-parton model as  
\begin{eqnarray}
 \frac{\Delta u+\Delta\bar u}{u+\bar u}&=&
	\frac{4}{15}\frac{g_1^p}{F_1^p}(4+\duratio)
	-\frac{1}{15}\frac{g_1^n}{F_1^n}(1+4\duratio)~;
 \nonumber\\
 \frac{\Delta d+\Delta\bar d}{d+\bar d}&=&
	\frac{4}{15}\frac{g_1^n}{F_1^n}(4+\frac{1}{\duratio})
	-\frac{1}{15}\frac{g_1^p}{F_1^p}(1+\frac{4}{\duratio}) ~,
\nonumber
\end{eqnarray}
where $\duratio=({d+\bar d})/({u+\bar u})$.  We performed a fit to the world $g_1^p/F_1^p$ 
data~\cite{thesis:zheng} and used $\duratio$ extracted from proton and 
deuteron structure function data~\cite{theory:duratio}.
Results for 
\renewcommand{\arraystretch}{1.25}
\begin{table}[h]
\begin{center}
\caption{Results for the polarized quark distributions.
The three errors are those due to the $g_1^n/F_1^n$ statistical error, $g_1^n/F_1^n$ 
systematic error and the uncertainties of $g_1^p/F_1^p$ and $\duratio$ fits.}
\label{tab:result_dqq}
\begin{ruledtabular}
\begin{tabular}{c|c|c}
 $x$  & $({\Delta u+\Delta\bar u})/({u+\bar u})$ & $({\Delta d+\Delta\bar d})/({d+\bar d})$  \\ \hline
$0.33$ & $0.565\pm 0.005^{+0.002}_{-0.002}~^{+0.025}_{-0.026}$ 
   &  $-0.274\pm 0.032^{+0.013}_{-0.013}~^{+0.010}_{-0.018}$ \\ 
$0.47$ & $0.664\pm 0.007^{+0.002}_{-0.002}~^{+0.060}_{-0.060}$ 
   &  $-0.291\pm 0.057^{+0.018}_{-0.018}~^{+0.032}_{-0.034}$ \\ 
$0.60$ & $0.737\pm 0.007^{+0.003}_{-0.003}~^{+0.116}_{-0.116}$ 
   &  $-0.324\pm 0.083^{+0.031}_{-0.031}~^{+0.085}_{-0.089}$ \\ 
\end{tabular}
\end{ruledtabular}
\end{center}
\end{table}
\renewcommand{\arraystretch}{1.0}
\begin{figure}[h]
  {\includegraphics[angle = 0, scale=0.7]{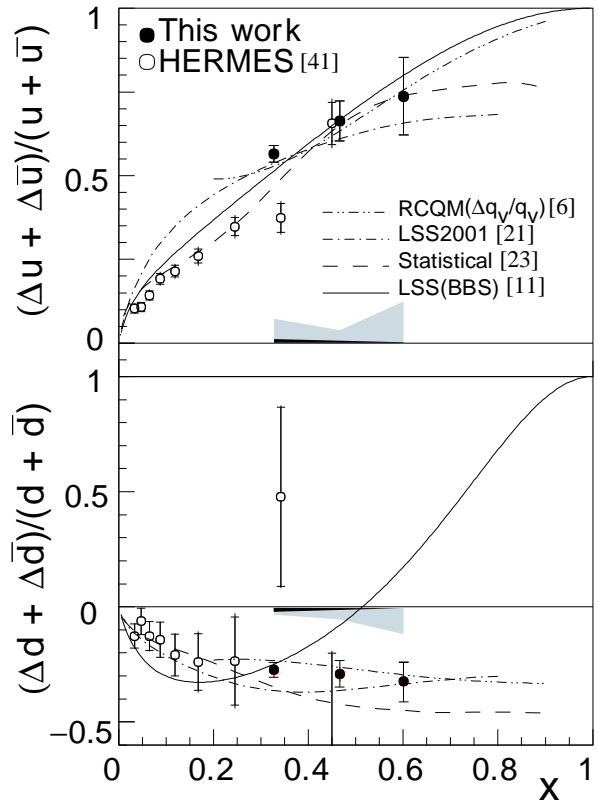}}
 \put(-112,277){\cite{data:hermes_dqq}}
 \put(-18,221.8){\cite{theory:cqm}}
 \put(-38,211.6){\cite{theory:lss2001}}
 \put(-36.5,202){\cite{theory:stat}}
 \put(-34,192.0){\cite{theory:lssbbs}}
 \caption{Results for {\fnz{$({\Delta u+\Delta\bar u})/({u+\bar u})$}} and 
{\fnz{$({\Delta d+\Delta\bar d})/({d+\bar d})$}} in the quark-parton model, 
compared with HERMES 
data~\cite{data:hermes_dqq}, the RCQM predictions~\cite{theory:cqm},
predictions from LSS~2001 NLO polarized parton densities~\cite{theory:lss2001},
the statistical model~\cite{theory:stat},
and pQCD-based predictions incorporating HHC~\cite{theory:lssbbs}.
The error bars of our data include
the uncertainties given in Table~\ref{tab:result_dqq}.
The dark-shaded error band on the horizontal axis shows
the uncertainty in the data due to neglecting $s$ and $\bar{s}$ contributions.
The light-shaded band shows the difference between $\Delta q_V/q_V$ and 
$(\Delta q+\Delta\bar q)/(q+\bar q)$ that needs to be applied to the data when comparing with
the RCQM calculation.}
\label{fig:result_dqq}
\end{figure}
$({\Delta u+\Delta\bar u})/({u+\bar u})$ and $({\Delta d+\Delta\bar d})/({d+\bar d})$ 
extracted from our $g_1^n/F_1^n$ data are listed in Table~\ref{tab:result_dqq}.

Figure~\ref{fig:result_dqq} shows our results along with HERMES data~\cite{data:hermes_dqq}.
The dark-shaded error band 
is the uncertainty due to neglecting the strangeness contributions.
To compare with the RCQM prediction which is given for valence quarks, 
the difference between $\Delta q_V/q_V$ and $ (\Delta q+\Delta\bar q)/(q+\bar q)$
was estimated and is shown as the light-shaded band.
Here $q_V$($\Delta q_V$) is the unpolarized (polarized) 
valence quark distribution for $u$ or $d$ quark.  Both errors were estimated 
using the CTEQ6M~\cite{theory:cteq} and MRST2001~\cite{theory:mrst} 
unpolarized parton distribution functions and
the positivity conditions that $\vert{\Delta q/q}\vert\leqslant~1$, 
$\vert{\Delta\bar q/\bar q}\vert\leqslant~1$ and $\vert\Delta q_V/q_V\vert\leqslant~1$.
Results shown in Fig.~\ref{fig:result_dqq} agree well with the predictions from 
RCQM~\cite{theory:cqm} and LSS 2001 NLO polarized parton densities~\cite{theory:lss2001}.  
The results agree reasonably well with the statistical model 
calculation~\cite{theory:stat} but do not agree with the predictions from LSS(BBS) 
parameterization~\cite{theory:lssbbs} based on hadron helicity conservation.

In summary, we have obtained precise data on the neutron spin asymmetry $A_1^n$ and the 
structure function ratio $g_1^n/F_1^n$ in the deep inelastic region at large $x$. 
Our data show a clear trend that $A_1^n$ becomes positive at large~$x$.
Combined with the world proton data, the polarized quark distributions 
$(\Delta u+\Delta \bar u)/(u+\bar u)$ and $(\Delta d+\Delta\bar d)/(d+\bar d)$ were extracted.
Our results agree with the LSS 2001 pQCD fit to the previous data and the trend 
agrees with the hyperfine-perturbed RCQM predictions. The new data do not agree 
with the prediction from pQCD-based hadron helicity conservation, 
which suggests that effects beyond leading order pQCD, such as the quark 
orbital angular momentum may play an important role in this kinematic region.
Extension of precision measurements of $A_1^n$ to higher $x$ and wider $Q^2$
range is planned with the future JLab 12 GeV energy upgrade.

We would like to thank the personnel of Jefferson Lab for their efforts which 
resulted in the successful completion of the experiment.
We thank S.~J.~Brodsky, L.~Gamberg, N.~Isgur, X.~Ji, E.~Leader, W.~Melnitchouk, 
D.~Stamenov, J.~Soffer, M.~Strikman, A.~Thomas, H.~Weigel 
and their collaborators for the theoretical support and helpful discussions. 
This work was supported by the Department of Energy (DOE), 
the National Science Foundation,
the Italian Istituto Nazionale di Fisica Nucleare,
the French Institut National de Physique Nucl\'{e}aire et de Physique des 
Particules,
the French Commissariat \`{a} l'\'{E}nergie Atomique
and the Jeffress Memorial Trust. 
The Southeastern Universities Research Association operates the Thomas 
Jefferson National Accelerator Facility for the DOE under contract DE-AC05-84ER40150.


\end{document}